\documentclass[floatfix, amsmath,amssymb,aps,pre,twocolumn, superscriptaddress]{revtex4-2}
\usepackage{mathtools,amsmath,graphicx,units}
\usepackage{dcolumn}
\usepackage{bm}
\usepackage{xcolor}
\usepackage{physics}
\usepackage{float}
\usepackage{epstopdf}
\usepackage{enumitem}
\usepackage[normalem]{ulem}
\usepackage[plainpages=false,pdfpagelabels,colorlinks=true,linkcolor=red,urlcolor=blue,citecolor=blue,pdftitle={Title},pdfauthor={},pdfdisplaydoctitle=true,pdfduplex=DuplexFlipLongEdge]{hyperref}
\usepackage[T1]{fontenc}

\usepackage[makeroom]{cancel}

\begin{document}

\title{ 
Thermalization Universality-Class Transition Induced by Anderson Localization
}
\author{Weihua Zhang}
\email{zhangwh2018@gmail.com}
\affiliation{Center for Theoretical Physics of Complex Systems, Institute for Basic Science, Daejeon 34126, Republic of Korea}
\affiliation{Lanzhou Center for Theoretical Physics and the Gansu Provincial Key Laboratory of Theoretical Physics, Lanzhou University, Lanzhou, Gansu 730000, China}
\author{Gabriel M.~Lando}
\email{gmlando@ibs.re.kr}
\affiliation{Center for Theoretical Physics of Complex Systems, Institute for Basic Science, Daejeon 34126, Republic of Korea}
\author{Barbara Dietz}
\email{bdietzp@gmail.com}
\author{Sergej Flach}
\email{sergejflach@googlemail.com}
\affiliation{Center for Theoretical Physics of Complex Systems, Institute for Basic Science, Daejeon 34126, Republic of Korea}
\affiliation{Basic Science Program, Korea University of Science and Technology (UST), Daejeon 34113, Republic of Korea}

\date{\today}
    
\begin{abstract}
We study the disorder-induced crossover between the two recently discovered thermalization slowing-down universality classes -- characterized by long- and short-range coupling -- in classical unitary circuits maps close to integrability. We compute Lyapunov spectra, which display qualitatively distinct features depending on whether the proximity to the integrable limit is short or long ranged. For sufficiently small nonlinearity, translationally invariant systems fall into the long-range class. Adding disorder to such a system triggers a transition to the short-range class -- implying a breaking of this invariance -- and in the very limit of vanishing non-linearity Anderson localization emerges. The crossover from long- to short-range class is attained by tuning the localization length, $\xi$, from $\xi \approx N$ to $\xi \ll N$, where $N$ is the system size. 
As a consequence, the Lyapunov spectrum becomes exponentially suppressed,
depending on how strongly its translational invariance is destroyed. 
We expect that this disorder-induced crossover will lead to prethermalized phases and, following quantization, to many-body localization.

\end{abstract}

\maketitle

Thermalization is a universal property of nonintegrable many-body systems, with characteristic timescales that diverge upon approaching integrability \cite{huang_statistical_1987,bel_weak_2005,rigol_breakdown_2009,danieli_intermittent_2017,danieli_dynamical_2019,mithun_dynamical_2019,mithun_fragile_2021}. In the near-integrable regime, one can interpret the system as a perturbation of an integrable one, with the perturbation's overall effect being to couple the action-angle variables of the unperturbed system in a nonlinear manner \cite{danieli_intermittent_2017,danieli_dynamical_2019,mithun_fragile_2021,mithun_dynamical_2019}. Thermalization slowing-down was shown to strongly depend on the perturbation's coupling range \cite{mithun_fragile_2021,danieli_dynamical_2019,wijn_lyapunov_2013}, which can be classified according to two universality classes: short-range network (SRN) and long-range network (LRN) \cite{danieli_dynamical_2019}.
\cite{goldfriend_equilibration_2019, ganapa_thermalization_2020,malishava_thermalization_2022,baldovin_statistical_2021}. 
To quantify these slowing-down processes, one can study finite time averages of observables. These, however need to be selected with care  \cite{malishava_lyapunov_2022,malishava_thermalization_2022,danieli_dynamical_2019} to ensure that one obtains proper ergodization timescales
\cite{goldfriend_equilibration_2019,ganapa_thermalization_2020,malishava_thermalization_2022, baldovin_statistical_2021}. 
The ambiguity in the choice of suitable observables led to the use of a novel method based on the Lyapunov spectrum (LS) to distinguish the two universality classes \cite{malishava_thermalization_2022}.
In this approach, which can also be employed to diagnose phase transitions \cite{dellago_lyapunov_1996,dellago_lyapunov_1997,bosetti_hadrien_what_2014,de_wijn_chaotic_2015}, significantly more information than the typical Lyapunov time (given by the inverse of the largest Lyapunov exponent) is available, since each Lyapunov exponent (LE) in the spectrum carries its own characteristic timescale.

The numerical challenges of dealing with weakly nonintegrable many-body Hamiltonian systems led to the study of one-dimensional unitary circuits maps, for which time-evolution is exact. This removes time discretization errors and allows for substantially larger evolution times, and thus for higher resolution in the LS \cite{malishava_lyapunov_2022,malishava_thermalization_2022}. The resulting universal scaling properties of the LS rendered possible an unambiguous identification of the different SRN and LRN universality classes. These were also observed in a recent study of multidimensional Josephson-junction networks across all possible lattice dimensions \cite{lando2023thermalization}. The predictive power of unitary circuits maps was thereby confirmed, and above all the LS has established itself as an invaluable tool in the study of the thermalization of many-body systems~\cite{malishava_lyapunov_2022}. 

Our objective is to gain insight into the interplay of disorder with nonintegrability, which might reveal a connection with the celebrated phenomenon of many-body localization~\cite{pal_many-body_2010}.
For this, we employ unitary circuits maps to investigate how the disorder impacts the two thermalization slowing-down universality classes. We use tailored disorder which leads to Anderson-localized states \cite{anderson_absence_1958} in the integrable limit of linear maps. The tunable localization length, $\xi$ is universal for all eigenmodes and is determined solely by the hopping-like parameter associated with the unitary circuits map \cite{vakulchyk_universal_2023}. We then demonstrate that the system's thermalization universality class changes from the LRN to SRN as the localization length is tuned from $\xi \approx N$ to $\xi \ll N$, where $N$ is the system size. Our findings intertwine the fields of many-body localization and thermalization of weakly non-integrable systems and open a new venue for connecting the slowing down of classical many-body dynamics in the presence of disorder and localization physics in quantum many-body systems like the ones proposed in Refs.~\cite{Chalker2018,Chalker2018a,Li2018,Nahum2018,Skinner2019,Li2019}.

\begin{figure}[h!]
    \centering
    \includegraphics[width=\linewidth]{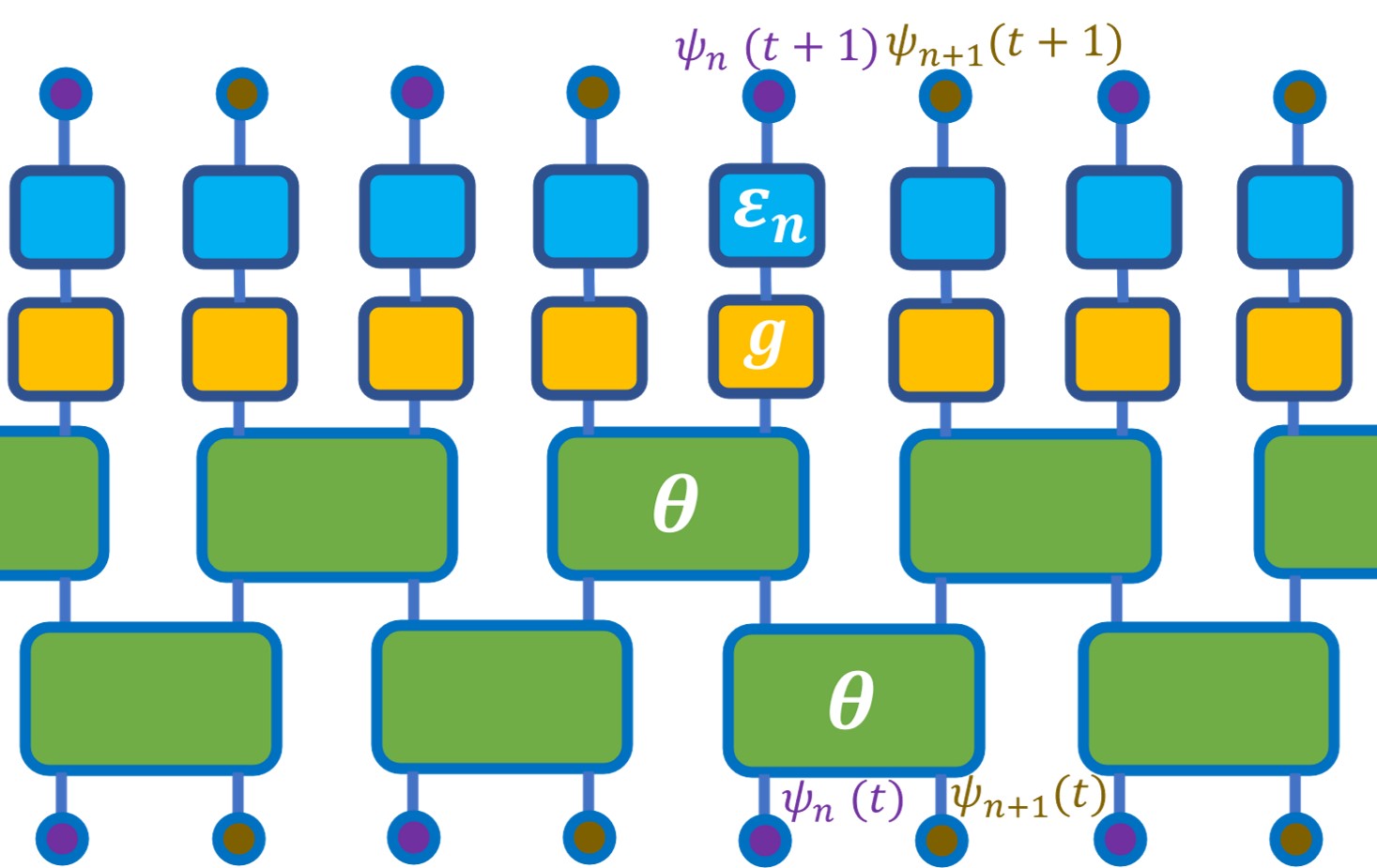}
    \caption{
        Schematic representation of the disordered unitary circuits map, where violet and brown circles indicate sites. Large green blocks represent unitary matrices $\hat C(\theta)$, small yellow blocks indicate local nonlinear unitary operators $\hat{G}(g)$, and small blue blocks indicate local disorder unitary operators $\hat{D}(\{\epsilon_n\})$. One time step contains four steps (unitary transformations). 
    }
    \label{fig1}
\end{figure}

We employ a modification of the classical unitary circuits maps introduced in \cite{malishava_lyapunov_2022, malishava_thermalization_2022}. These consist of a one-dimensional periodic chain of $N$ complex numbers, $\psi_n$, with $n=1,2,\dots, N$ denoting the sites (we assume that $N$ is even). An initial vector 
$\vec{\Psi}$ with these complex scalar components $\psi_n$ is evolved by applying iteratively the unitary map
\begin{equation}\label{eq:st ep}
    \widehat{U} = \widehat{D}\widehat{G}\widehat{C}^{(\mathrm{even})}\widehat{C}^{(\mathrm{odd})} \, .
\end{equation}
Here, we interpret the number of iterations as a discrete-time,  i.e.,~$\vec{\Psi}(t+1) = \widehat{U}\vec{\Psi}(t)$. A pictorial description of $\widehat{U}$ is provided in Fig.~\ref{fig1}. 
The operators $\widehat{C}^{(\mathrm{even/odd})}$
are linear transformations. In matrix representation they are block diagonal with each block consisting of a $2\times 2$ unitary matrix $\widehat{C}_{n,n+1}$ coupling the components $\psi_{n}(t)$ and $\psi_{n+1}(t)$, parameterized by a hopping-like angle parameter $\theta$,
\begin{equation}\label{eq:Cn}
    \widehat{C}_{n,n+1}
    \begin{pmatrix}
       \psi_n(t) \\ 
        \psi_{n+1}(t) 
    \end{pmatrix}
    =
    \begin{pmatrix}
        \cos \theta & \sin \theta \\ 
        -\sin \theta & \cos \theta 
    \end{pmatrix}
    \begin{pmatrix}
       \psi_n(t) \\ 
        \psi_{n+1}(t) 
    \end{pmatrix}
    \, .
\end{equation}
All remaining matrix elements are zero and, due to periodicity, $\psi_{N+1} = \psi_1$. 
Successive applications of 
$\widehat{C}^{(\mathrm{odd})}$ and $\widehat{C}^{(\mathrm{even})}$
intertwine neighbors to the left and to the right of site $n$ (green boxes in Fig.~\ref{fig1}). In matrix representation the operator $\widehat{G}$ corresponds to a diagonal matrix whose elements $\widehat{G}_{n,n}$ are nonlinear on-site potentials acting as
\begin{equation}
\widehat{G}_{n,n} \psi_n(t)  = e^{i g \, \vert \psi_n \vert^2} \psi_n(t) \, .
\end{equation}
Similar to an anharmonic oscillator whose frequency depends on its level of excitation, the imposed phase shift (during one given time step) depends on the amplitude $|\psi_n|$.
This term is responsible for the chaoticity of the dynamics (yellow boxes). Finally, $\widehat{D}$ acts on a given site, that is, in matrix representation again is a diagonal matrix with elements 
\begin{equation}
\widehat{D}_{n,n} \psi_n(t)  = e^{i \epsilon_n } \psi_n(t) \, ,
\end{equation}
where the $\epsilon_n$ are site-dependent disorder potentials that are uniformly distributed in $[-\pi, \pi]$ (blue boxes). 
The unitarity of $\widehat{U}$ implies that the total squared-norm $\vert \vec{\Psi}(t) \vert^2$ is a conserved quantity. Accordingly, in order to allow all possible typical scenarios for the temporal behavior with equal probability, we chose the initial values of the squared moduli of the rescaled components $\eta_n =N\vert\psi_n\vert^2$ uniformly spread over the $N$ sphere with the joint-probability distribution $P(\{\eta_n\})\propto\delta\left(N-\sum_{n=1}^N\eta_n\right)$. This yields for the probability distribution of the $\eta_n$ $P(\eta)\propto (1-1/N)(1-\eta/N)^{N-2}\xrightarrow{N\gg 1}e^{-\eta}$. For the computation of the LS, $\Lambda$, comprising the LEs, $\Lambda_i$, with $\Lambda_1\geq \Lambda_2\cdots\geq\Lambda_{N}$, we follow the calculation procedure of Ref.~\cite{malishava_lyapunov_2022}, outlined in more detail in the appnedix. Note that we are considering the part of the LS which is composed of non-negative LEs in the spectrum. We can do so because, similarly to time-independent Hamiltonian systems, the $N$ positive LEs come in pairs with negative ones of the same modulus,  $\Lambda_i = - \Lambda_{2N-i+1}$ \cite{malishava_lyapunov_2022}. The final simulation time is denoted by $T_s$.

In the absence of hopping, achieved by setting $\theta=0$, the sites decouple and the squared-norms $\vert \psi_n \vert^2$ are conserved. This results in $N$ functionally independent conserved quantities. Since the system has $N$ degrees of freedom implying a $2N$ dimensional phase space, this limit is integrable. The squared-norms then provide a discrete analog to the actions in continuous Hamiltonian systems when $\theta = 0$ \cite{malishava_lyapunov_2022, malishava_thermalization_2022}. A second integrable limit is realized by setting $g = 0$. In this case, the sites remain coupled but there is no source of nonlinearity, and the squared-norms of the corresponding normal modes,
\emph{i.e.}~of the eigenvectors of $\hat{U}(g=0)$, are conserved \cite{malishava_lyapunov_2022, malishava_thermalization_2022}.

\begin{figure}[htbp]
\centering
\includegraphics[width=0.9\linewidth]{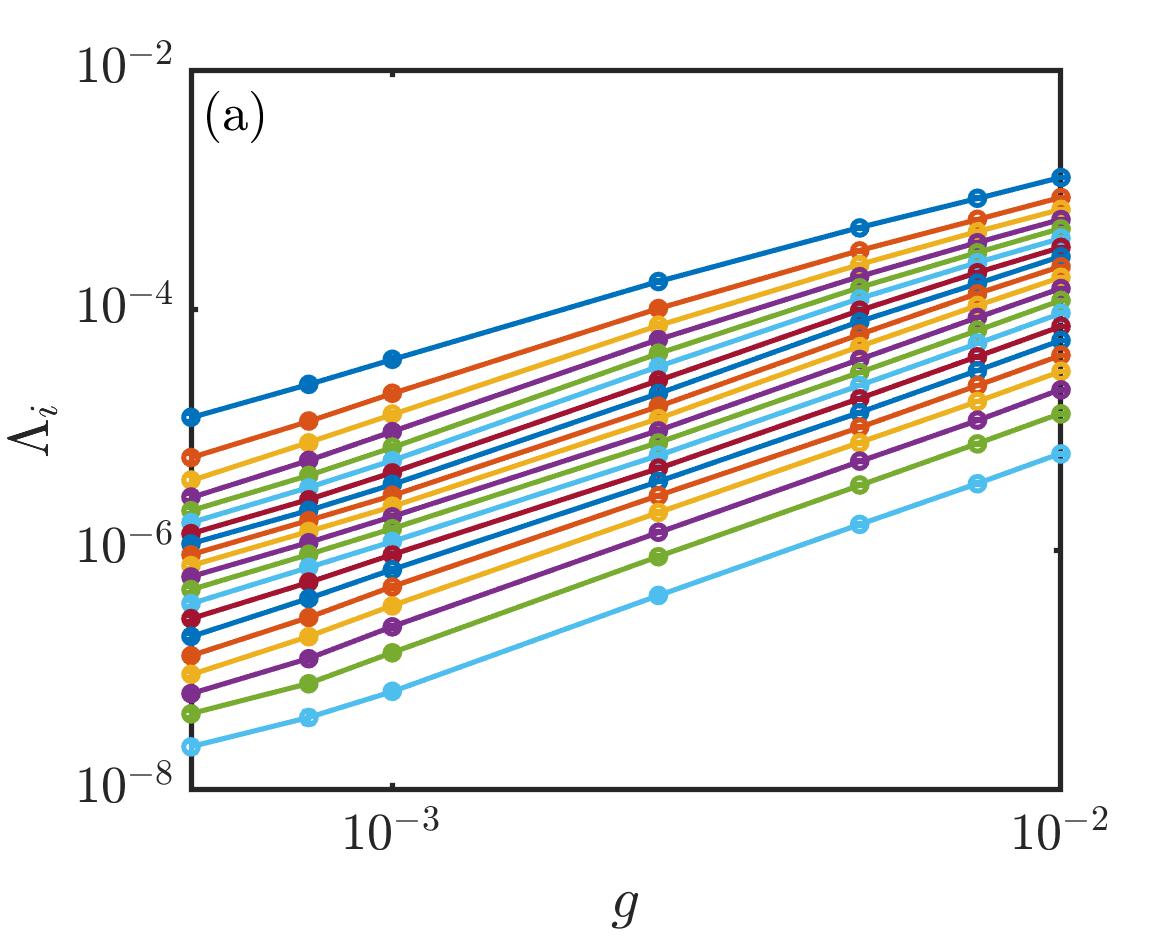}
\includegraphics[width=0.9\linewidth]{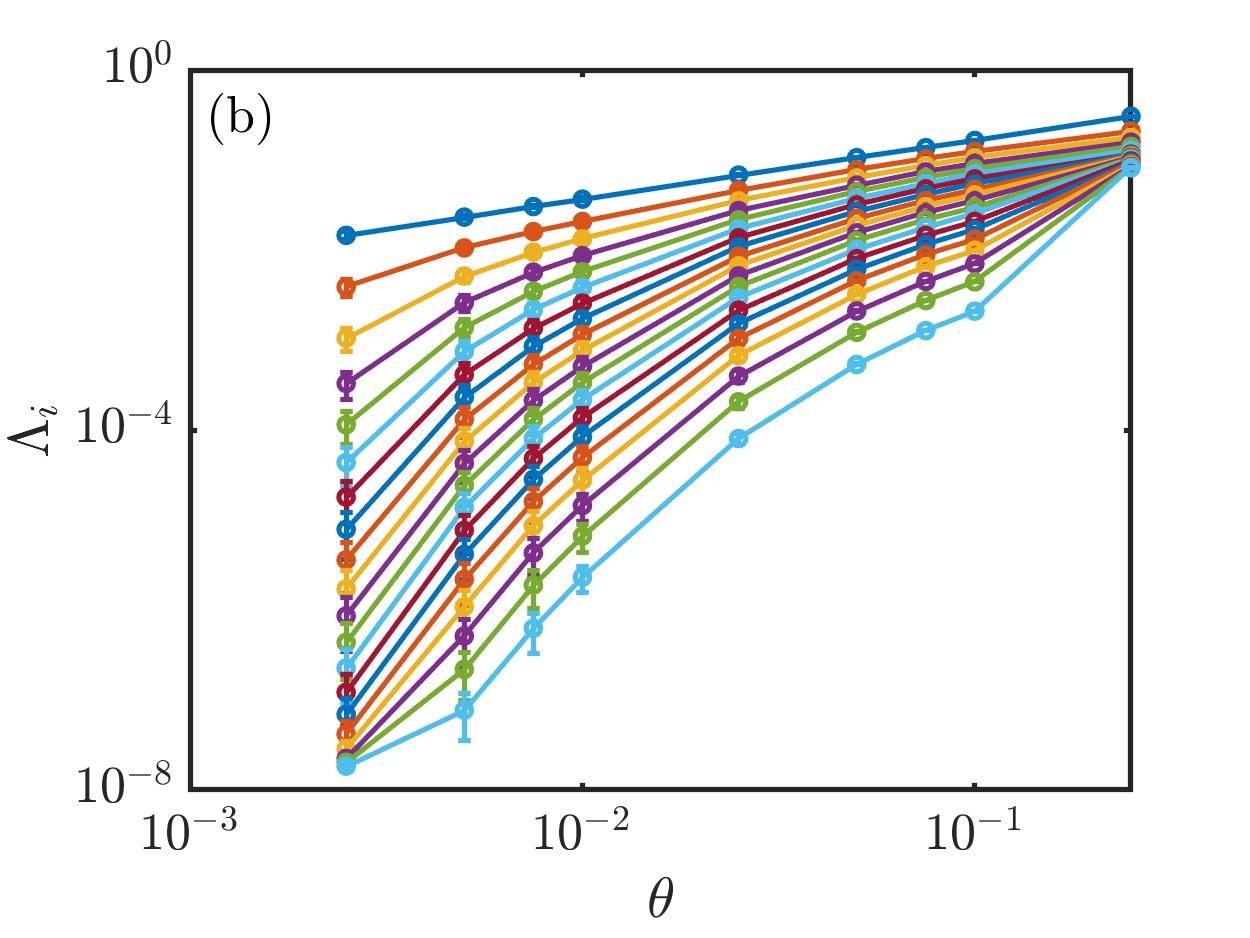}
\caption{
Ordered case.
Lyapunov spectra $\Lambda_i$ of the ordered unitary circuits map versus  (a) $g$ in the LRN regime ($\theta=1.13$, $T_s=10^8$), and (b) $\theta$ in the SRN regime ($g=1$, $T_s=10^9$).
The error bars show the standard deviation $\sigma_g$ obtained from an ensemble of 12 different trajectories. Here $N=200$.
}
\label{fig2}
\end{figure}

\begin{figure}[htbp]
    \centering
\includegraphics[width=1\linewidth]{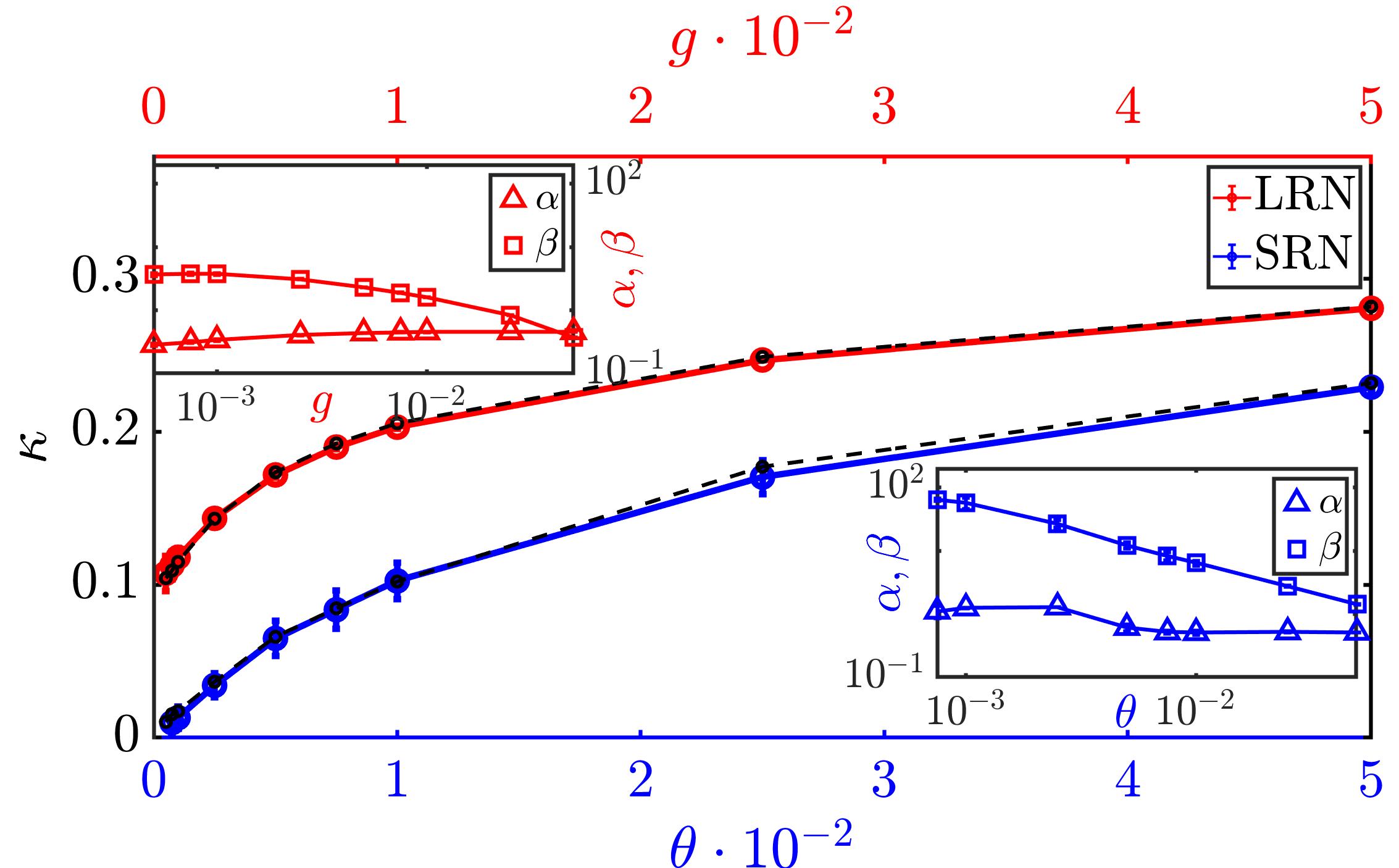}
    \caption{
    Ordered case. Rescaled KS entropy $\kappa$ in the LRN regime (red circles, top, $\theta=1.13$) versus $g$, and SRN regime (blue circles, bottom, $g=1$) versus $\theta$. Solid lines connect the data points and guide the eye. Here $T_s=10^8$, $N=200$. The insets show the fit coefficients $\alpha$ (triangles) and $\beta$ (squares) of Eq.~(\ref{eq:fit}) in the LRN (top left) and SRN (bottom right) regimes. The error bars represent the standard deviation deduced from an ensemble of 12 trajectories. The dashed curves in the main plot show the integral of Eq.~(\ref{eq:fit}) evaluated with the coefficients from the fits.}
\label{fig3}
\end{figure}

Let us consider first the disorder-free case. When approaching the decoupled integrable limit, i.e.,~when $\theta \ll 1$, each action is only weakly coupled to its nearest neighbors, resulting in a SRN \cite{malishava_lyapunov_2022, malishava_thermalization_2022}.
In the regime $g \ll 1$, on the other hand, all normal mode actions are weakly coupled to each other, such that all-to-all \textit{linear} interactions persist and implicate a LRN \cite{malishava_lyapunov_2022, malishava_thermalization_2022}.
Furthermore, it was demonstrated that, in the absence of disorder, the Lyapunov spectra for LRN and SRN systems 
behave significantly differently as the integrable limit is approached: While most of the LEs in the LRN class are of the same order of magnitude at any distance from the integrable limit, the LEs in the SRN class are damped and end up spanning several orders of magnitude, with the span increasing upon approaching the integrable limit (see Fig.~\ref{fig2}).
We derived an ansatz for the rescaled Lyapunov spectrum $\bar{\Lambda}$ of the LRN and SRN. This rescaled spectrum is obtained by dividing the original LEs by the maximal LE (mLE), $\bar\Lambda_i=\Lambda_i/\Lambda_1$. Starting point of the derivation is the analytical result for the number of Lyapunov exponents $N(\Lambda)$ below $\Lambda=\Lambda_i$, given by $N(\Lambda_i)=N+1-i,\, i=1,2,\dots,N$. Away from the limiting values, $0<\bar\Lambda_i<1$, it is well described by the integrated Wigner semicircle law~\cite{Mehta1990,Gur-Ari2016,Hanada2016} yielding an inverse semicircle law for $\bar\Lambda$ versus $\rho_i=i/N$. The damping rate of the spectrum in the SRN regime, on the other hand, is exponential. An approximation of the resulting expression, which for $\rho>0$ reflects these features, is provided by the ansatz
\begin{equation}\label{eq:fit}
    \overline{\Lambda(\rho)} =(1-\rho^{\alpha})e^{-\beta \rho^\gamma}.
\end{equation}
The parameters $\alpha,\beta,\gamma$ are determined from the fit of this ansatz to the rescaled LS. Yet, the qualitative behavior of $\overline{\Lambda (\rho)}$ is already well captured with $\gamma=1$~\cite{lando2023thermalization}. 
In addition, we compute the rescaled Kolmogorov--Sinai (KS) entropy $\kappa=\frac{1}{{N}-1}\sum_{i=2}^{{N}} \bar{\Lambda}_i = \int_0^1 \bar{\Lambda} \, 
\textrm{d}\rho$, which is a very useful quantifier of the different regimes. In the SRN regime, $\kappa$ tends to 0 when approaching the integrable limit, while in the LRN regime, it saturates at some value, as shown in Fig.~\ref{fig3}. 
The insets of Fig.~\ref{fig3} show that the fitting parameter $\alpha$ does not vary significantly for all considered cases. In contrast, 
the exponent $\beta$ grows with a power law in the SRN upon lowering $\theta$, while it saturates similarly to $\alpha$ in the LRN with decreasing $g$. 
As a consequence, the rescaled LS $\bar{\Lambda}(\rho)$ saturates on some analytic invariant curve in the LRN for $g \rightarrow 0$, and knowing one LE (e.g. the mLE) results in knowing all others as well, whereas the SRN is characterized by an exponential damping of LEs compared to the mLE. 

Our next objective is to get insight into the effect of disorder on the LRN with the aim to realize a SRN for weak, but nonzero nonlinearity $g \ll 1$. For this
we now move to the disordered case to connect to Anderson localization \cite{anderson_absence_1958,hamza_dynamical_2009, lagendijk_fifty_2009}.
For $g=0$, all normal mode eigenstates of $\hat{U}$ are exponentially localized~\cite{vakulchyk_universal_2023}. The localization length, $\xi$  depends only on the hopping-like parameter $\theta$,
\begin{equation}\label{Eq:loc_theta}
    \frac{2}{\xi}=|\ln(|\sin\theta|)|\;,\; \xi(\theta\rightarrow 0) \rightarrow 0\;,\; \xi(\theta \rightarrow \frac{\pi}{2}) \rightarrow \infty. 
\end{equation}
Let us expand an arbitrary $\vec{\Psi}(t)$ in the eigenmode basis at $g=0$, $\hat{U}(g=0) \vec{\Psi}_k = e^{i\omega_k} \vec{\Psi}_k$, namely $\vec{\Psi}(t)=\sum_k c_k(t) \vec{\Psi}_k$. Here, the eigenmodes $\vec{\Psi}_k$ are Anderson-localized with components $|\psi_k^n| \sim e^{-|n|/\xi}$.  The quasienergies $\omega_k$ are real, and the expansion coefficients $c_k(t)$ are complex. The actions $\{|c_k|^2\}$ are the constants of motion. Away from the integrable limit, \emph{i.e.}~for $0 < g \ll 1$, the expansion coefficients are coupled as
\begin{align}\label{Eq:coupling}
    &c_k(t+1)=e^{i\omega_{k}}c_k(t) + \\
    &\quad ig\sum_{k_1,k_2,k_3} 
    e^{i(\omega_{k_1}+\omega_{k_2}-\omega_{k_3})}I_{k,k_1,k_2,k_3}c_{k_1}(t)c_{k_2}(t)c^*_{k_3}(t), \notag
\end{align}
where $I_{k,k_1,k_2,k_3}$ is an overlap integral
$I_{k,k_1,k_2,k_3} \sim \sum_{n}\psi_{k_1}^{n}\psi_{k_2}^{n}(\psi_{k_3}^{n})^* (\psi_{k}^{n})^*$.
For small $g$ and $|\frac{\pi}{2}-\theta|$ the length $\xi$ becomes larger than the system size, implying that the normal modes extend over the entire system, and Eq.~(\ref{Eq:coupling}) turns into a LRN with essentially all-to-all action couplings. For small $g$ and $\theta \ll 1$ the length $\xi$ tends to zero, the eigenmodes are strongly localized, and Eq.~(\ref{Eq:coupling}) turns into a SRN with essentially nearest neighbor action couplings. 

\begin{figure}[htbp]
\centering
\includegraphics[width=0.9\linewidth]{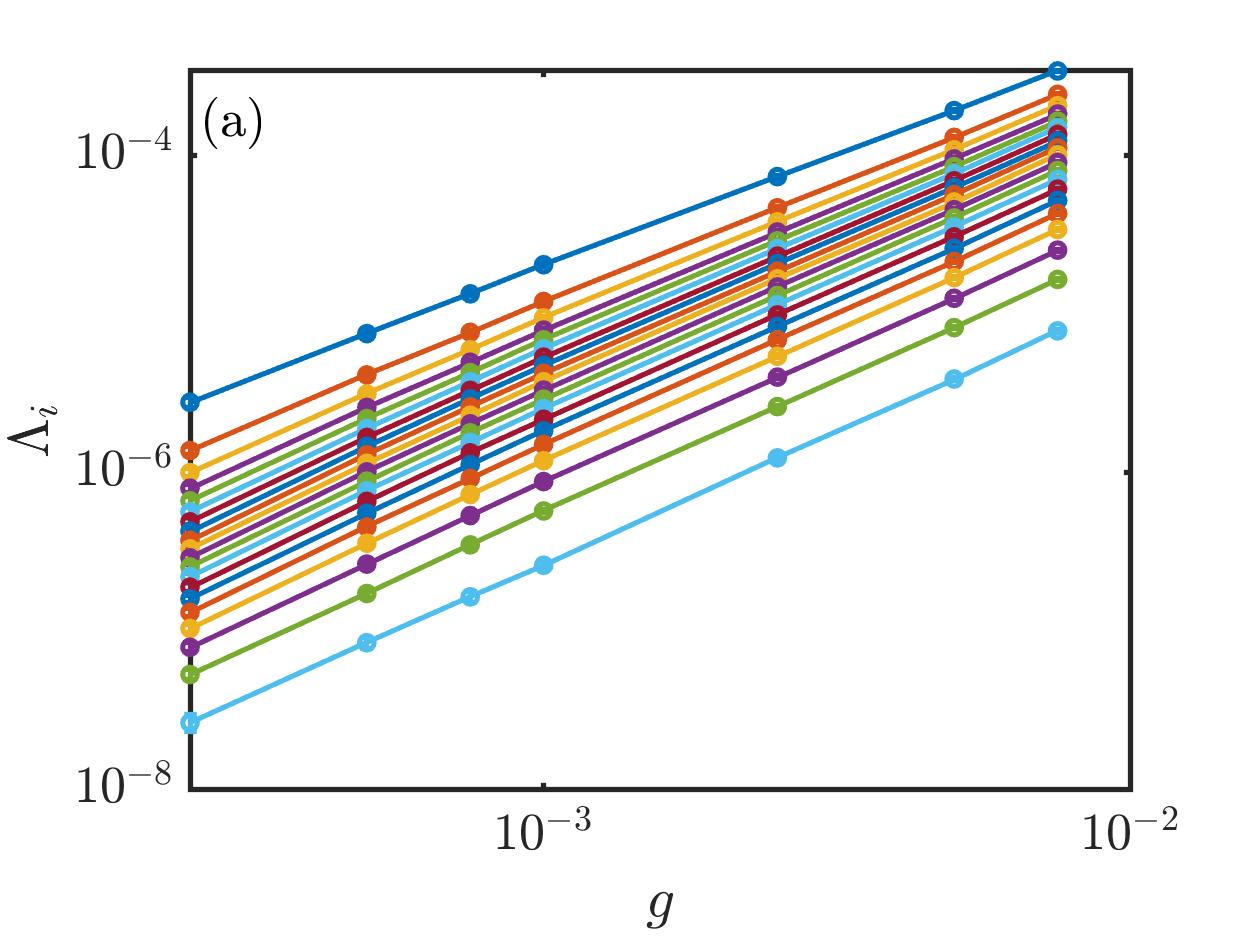}
\includegraphics[width=0.9\linewidth]{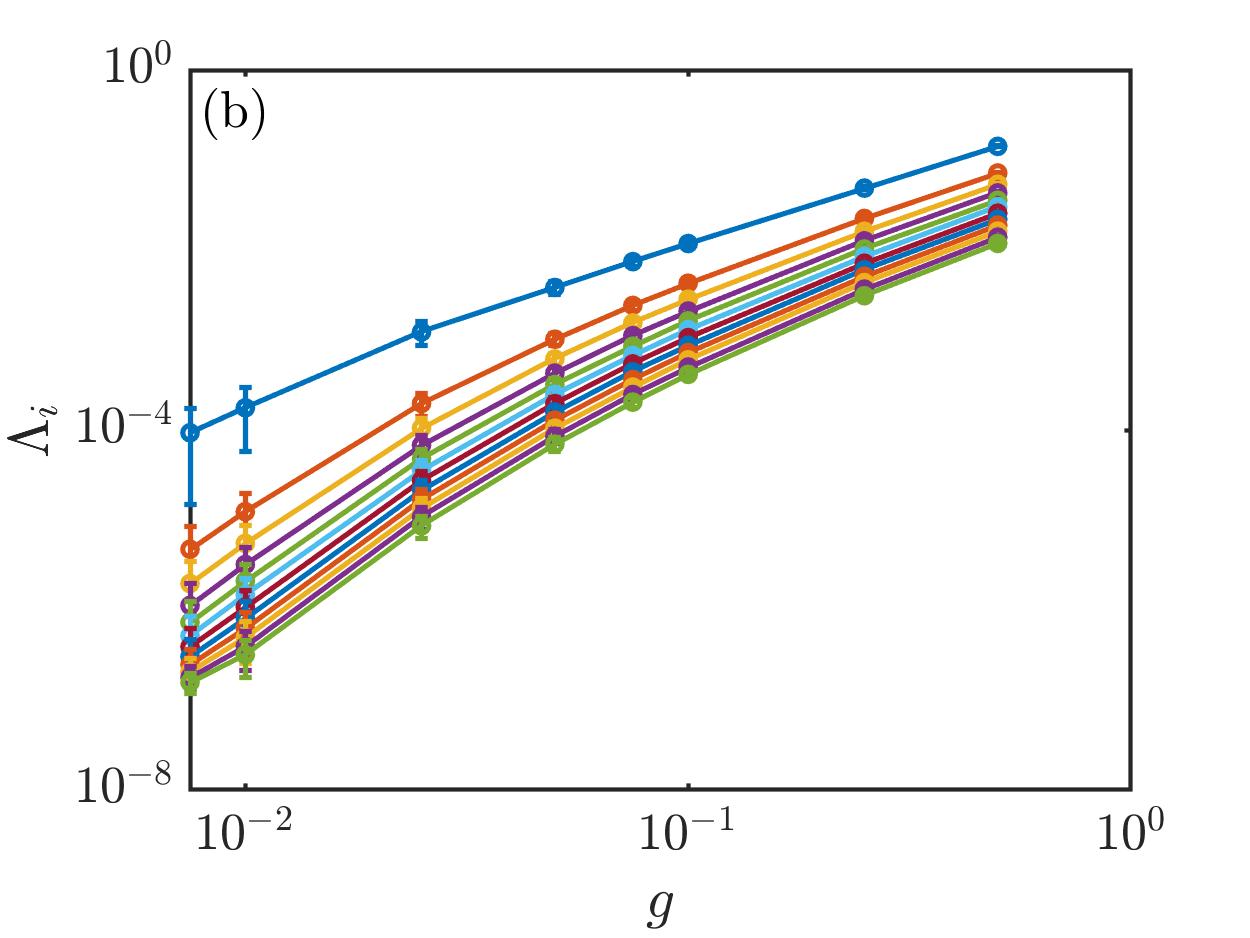}
\caption{
Disordered case. $\Lambda_i$ versus $g$, $T_s=10^8$, $N=200$.
The error bars show the standard deviation $\sigma_g$ for an ensemble of 100 trajectories.
(a) LRN with $\theta=1.26$ and $\xi \approx 40$. 
(b) SRN with $\theta=0.38$ and $\xi \approx 2$. 
}
\label{fig4}
\end{figure}

We, therefore, predict that the
universality class of a finite disordered system will depend on the ratio $\xi/N$. In Fig.~\ref{fig4} we show the LS dependence on $g$ for (a) $\xi/N=0.2$ and (b) $\xi/N=0.01$. 
We clearly observe that case (a) displays the typical LRN behavior observed in Fig.~\ref{fig2} with all LEs  being proportional to each other. Instead for case (b) the LEs widen in a fan-like way, indicating SRN features.

To gain further insight, we extrapolate the rescaled LS down to $g=0$ as a function of $\theta$ and fit the ansatz Eq.~(\ref{eq:fit}) to the outcome; see the appendix. 
The results are summarized in Fig.~\ref{fig5} and for $\alpha$ in Fig. A5 of appendix. Panel (a) shows the dependence $\xi(\theta)$ for convenience. 
Figure~\ref{fig5}(b) exhibits the dependence of the exponent $\beta$ on $\theta$  (filled squares), i.e. on the localization length $\xi$ with a growth over several decades as expected when crossing over from a LRN to a SRN class. 
In Fig.~\ref{fig5}(c) the dependence of the asymptotic rescaled KS entropy $\kappa_a$ versus $\theta$ is plotted (filled circles) showcasing the vanishing of $\kappa_a$ for small $\theta$. For reference, we plot in panels (b) and (c) the corresponding results for the ordered case using open symbols. In this case no sizable change of $\beta(\theta)$ and $\kappa_a(\theta)$ is observed. In addition to inducing the transition from a LRN to a SRN by varying the localization length $\xi$, this can be achieved by increasing the system size $N$. To demonstrate this, we have calculated the quantity $\kappa_a$ for different system sizes, all with the same localization length $\xi=2$, exhibited in Fig.~A6 of the appendix, yielding that $\kappa_a$ indeed decreases with increasing system size $N$.
\begin{figure}[htbp]
\centering
\includegraphics[width=0.9\linewidth]{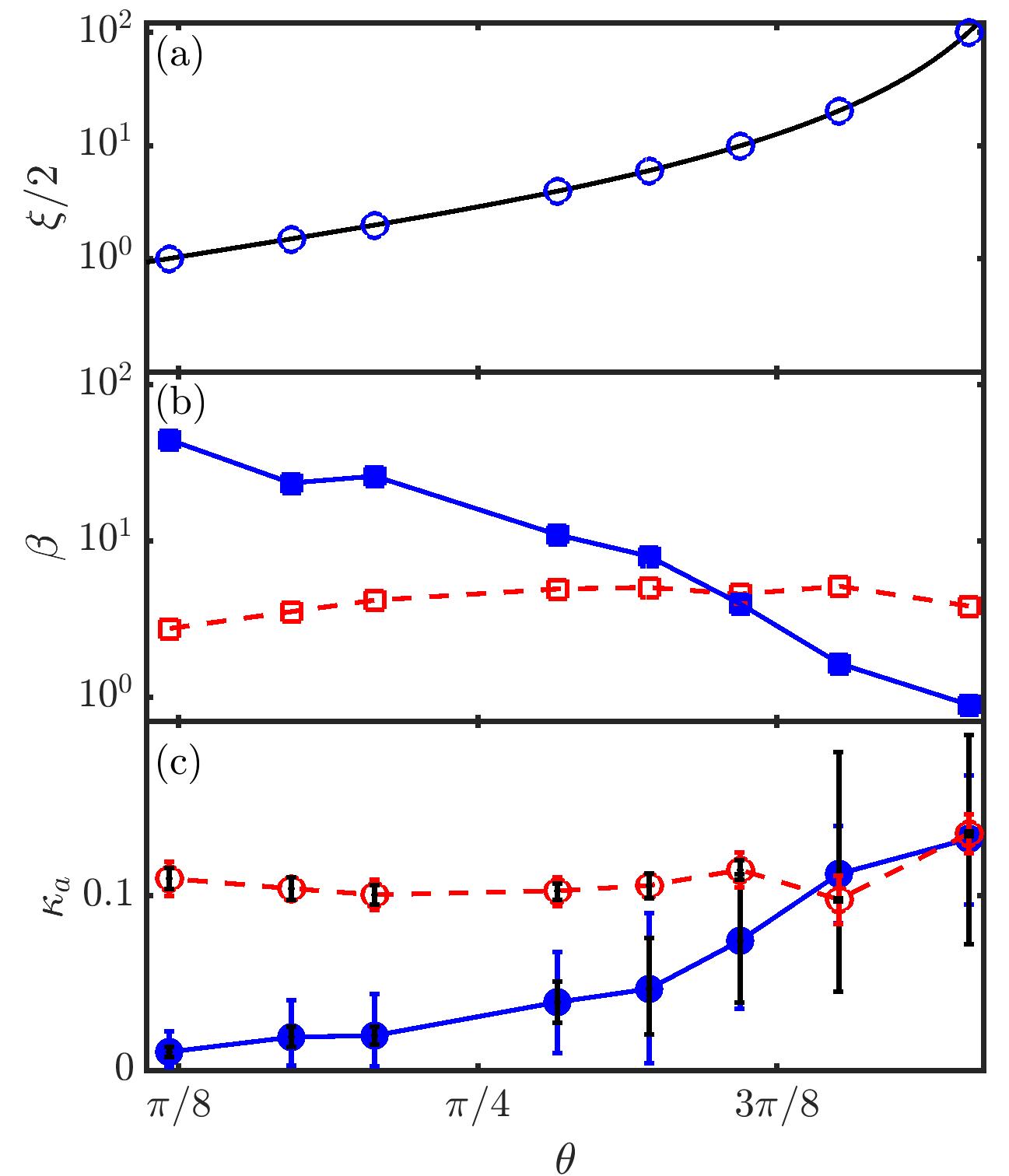}
\caption{
Disordered case after extrapolating down to $g=0$. Ordered case results are added for comparison.
(a) Localization length $\xi$ versus $\theta$. The black line is the analytical result of Eq.~(\ref{Eq:loc_theta}), and the blue circles mark the $\theta$ values used for the computations. 
(b) Fit coefficient $\beta$ of the asymptotic rescaled LS $\bar{\Lambda}(\rho; g\rightarrow 0)$ versus $\theta$, filled blue squares. Open red squares are obtained from the corresponding ordered case data; see appendix.
(c) Asymptotic rescaled KS entropy $\kappa_a$ versus $\theta$, filled blue circles. Open red circles are obtained from the corresponding ordered case data; see appendix.
The error bars represent the time ($\sigma_t$, black) and ensemble ($\sigma_g$, blue and red) standard deviations; see appendix.
Lines guide the eye.
}
\label{fig5}
\end{figure}

The Lyapunov spectrum (or better the inverse of its Lyapunov exponents) captures the timescale of thermalization. These times diverge in a qualitatively different way upon approaching integrable limits for short and long range network classes. Previous studies used the limit of weak lattice coupling to realize a SRN, while weak nonlinearities resulted in a LRN. Here, we show that the addition of disorder allows us to realize a SRN in the limit of weak nonlinearities, which in many situations correspond to weak two-body interactions. 
To achieve our goals we compute the rescaled Lyapunov spectrum and fit the ansatz Eq.~(\ref{eq:fit}) to it to extract an exponent whose divergence signals the presence of a SRN, and measure the rescaled KS entropy of the rescaled LS as another highly useful quantifier to tell SRN and LRN regimes apart. We also extrapolate the rescaled LS down to vanishing nonlinearity strength. These procedures allow us to unambiguously identify the crossover from a long range network to a short range network in a disordered system upon reducing the localization length. 

Let us discuss and speculate about some consequences of the observed crossover. Approaching the integrable limit in the regime of a short range network implies that at any time the dynamics of the system is mostly regular, with rare local spots of nonlinear resonances leading to weak chaos (which is probably associated to the largest Lyapunov exponent). The density of these rare spots diminishes the closer the system is tuned to the integrable limit. Large but finite systems are therefore expected to display prethermalization features, where certain parts of the system show thermal properties which will fluctuate from part to part. Additional quantization of the considered systems might lead to a suppression of the chaotic resonances and the prethermalization, and ultimately to many-body localization related features. It remains to be studied whether the well defined short range network regimes of the studied classical systems will indeed result in many-body localization for the corresponding quantum systems, at variance to the long range network classes.

Our studies were confined to finite system sizes. We expect that the thermodynamic limit $N\to \infty$  will result in a short range network class for all cases of finite localization length; see Sect.~VII of the appendix. It remains to be studied whether large but finite values of the localization length will or will not result in qualitative changes of the Lyapunov spectrum scaling in the limit of weak nonlinearities and infinite system size.

\begin{acknowledgements} 
This work was supported by the Institute for Basic Science in Korea through the project IBS-R024-D1. WZ acknowledges financial support from the China Scholarship Council (No. CSC-202106180044). WZ also acknowledges financial support from the NSF of China under grant Nos. 11775100, 12247101, and 11961131009. We thank Tilen \v{C}ade\v{z} and Alexei Andreanov for fruitful discussions.
\end{acknowledgements}

\appendix

\begin{appendix}
\renewcommand\thefigure{\thesection.\arabic{figure}}
\renewcommand{\theequation}{A\arabic{equation}}
\renewcommand{\thefigure}{A\arabic{figure}}
\setcounter{figure}{0}

\section{Dynamic evolution}
For the sake of simplicity, we use the notation $\{\psi_{2n}(t), \psi_{2n+1}(t)\}^T$ to represent the states of two sites within a single unit cell (depicted as violet and brown circles in Fig.~1 of the main text). The time evolution of these states is governed by the equations \cite{malishava_lyapunov_2022,malishava_thermalization_2022}:

\begin{align}\label{Eq:Evolutionnonlinear}
&\psi_{n}(t+1)=e^{i\epsilon_{n}}e^{i(g|f_{n}[\Vec{\Psi}(t)]|^2)}f_{n}[\Vec{\Psi}(t)], \\
\end{align}

with
\begin{equation}\label{Eq:Evolutionlinear}
\scalebox{0.9}{$
\begin{gathered}
f_{2n}[\Vec{\Psi}(t)]=\mathcal{C}^2\psi_{2n}(t)-\mathcal{C}\mathcal{S}\psi_{2n-1}(t)+\mathcal{S}^2\psi_{2n-2}(t)+\mathcal{C}\mathcal{S}\psi_{2n+1}(t), \\
f_{2n-1}[\Vec{\Psi}(t)]=\mathcal{C}^2\psi_{2n-1}(t)-\mathcal{C}\mathcal{S}\psi_{2n-2}(t)+\mathcal{S}^2\psi_{2n+1}(t)+\mathcal{C}\mathcal{S}\psi_{2n}(t),
\end{gathered}
$}
\end{equation}
where $\mathcal{C}=\cos\theta$ and $\mathcal{S}=\sin\theta$.

\section{Initial conditions}
In all simulations, we initialize the system with initial conditions ${|\psi_n|e^{i\phi_n}},\, n=1,2,\dots N$ for the $n$th component of $\Vec{\Psi}(0)$, where the phases $\phi_n$ are uncorrelated and uniformly randomly distributed in the range $[-\pi, \pi]$, and the rescaled squared amplitudes $\eta_n=N|\psi_n|^2$ follow an exponential distribution $p(\eta)\propto e^{-{\eta/\langle\eta\rangle}}$. The average value of the square-norm of $\psi_n$ is normalized to $\langle\eta\rangle=\frac{1}{2}$ which is consistent with the approach used 
in Refs.~\cite{malishava_lyapunov_2022,malishava_thermalization_2022}.

The strengths of disorder, denoted by $\epsilon_n$, are also uncorrelated and uniformly randomly distributed in the interval $[-\pi, \pi]$, following the procedure described in Ref.~\cite{vakulchyk_universal_2023}.

\section{Lyapunov Spectrum Calculation}
\begin{figure}
\centering
\includegraphics[width=0.9\linewidth]{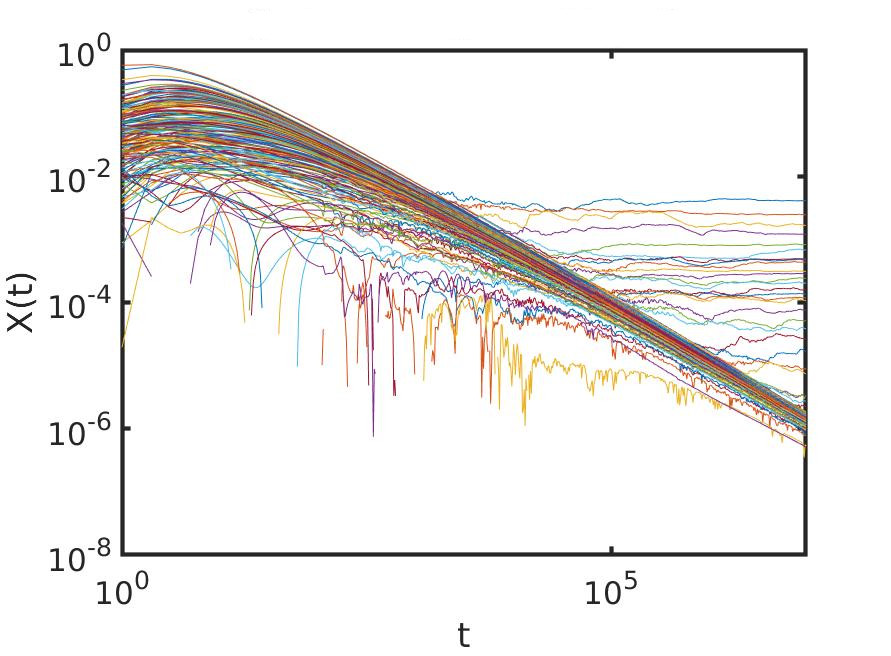}
\caption{Evolution of positive transient LS with $\theta=0.00075$ and $g=1$. $N=100$, $T_s=10^7$.}
\label{figS1}
\end{figure}
The Lyapunov exponent measures the exponential growth rate of an infinitesimal distance $w(t)$ from the initial trajectory. It is defined as $\Lambda = \lim_{t\to \infty} \frac{1}{t} \ln \frac{w(t)}{w(0)}$ \cite{skokos_lyapunov_2010}. To compute the positive LS, we introduce $N$ orthogonal perturbations $\Vec{w}(t)$ to trajectories $\Vec{x}(t)$, where $N$ is equal to the degree of freedom of the system and the state $\Vec{\Psi}(t) = \Vec{x}(t) + \Vec{w}(t)$ follows the equations of motion Eq.~(\ref{Eq:Evolutionnonlinear}).

Because of the linearity property $f_n[\Vec{\Psi}(t)] = f_n[\Vec{x}(t)] + f_n[\Vec{w}(t)]$, we can expand the norm in the exponential term in Eq.~(\ref{Eq:Evolutionnonlinear}). We only keep the first order of $\Vec{w}$,
\begin{align*}
    &|f_n[\Vec{\Psi}(t)]|^2 =|f_n[\Vec{x} (t)]+f_n[\Vec{w}(t)]|^2\\ 
    &=|f_n[\Vec{x}(t)]|^2+|f_n[\Vec{w}(t)]|^2+2\Re{f_n[\Vec{w}(t)]f_n[\Vec{x}(t)]^*}\\
    &\approx |f_n[\Vec{x}(t)]|^2+\Re_n
\end{align*}
where $\Re_n=2\Re{f_n[\Vec{w}(t)]f_n[\Vec{x}(t)]^*}$.

Expanding the nonlinear part in a Taylor series and retaining only the first order of $\Vec{w}(t)$, we obtain
\begin{align*}
    e^{ig|f_n[\Vec{\Psi}(t)]|^2}
    &=e^{i|f_n[\Vec{x}(t)]|^2}e^{ig\Re_n}\\
    &\approx e^{i|f_n[\Vec{x}(t)]|^2}(1+ig\Re_n).
\end{align*}
Subtracting the contribution from the linear trajectories yields the growth of the perturbation
\begin{align*}
    &w_n(t+1)
    =\psi_n(t+1)-x_n(t+1)\\
    &=e^{i\epsilon_n}{e^{i(g|f_n[\Vec{x}(t)]|^2)}}\{f_n[\Vec{w}(t)]+ig\Re_n f_n[\Vec{x}(t)]\}.
\end{align*}

Fig.~\ref{figS1} exemplifies the evolution of the positive transient LS for one specific case.

\section{Saturation criterion and Asymptotic curve of rescaled Lyapunov spectrum}

\subsection{Saturation criterion of Lyapunov spectrum}
As we approach an integrable limit, the divergent timescales become inversely proportional to the LS. However, due to computational limitations, we are required to terminate our evolution at a specific number of time steps, typically ranging from $10^8$ to $10^9$ in this study. Despite this limitation, even at these time steps, certain parts of the LS may not have reached saturation. To quantify the saturation level, we calculate the standard deviation $\sigma_t$ of the data between the final time, $10^{n_s}$, and its previous ``generation'', $10^{n_s-1}$. This interval typically covers more than 90\% of the total evolution time.

To control the accuracy of our simulations, we utilize the parameter $L = \frac{\sigma_t}{\Lambda_{\text{max}}}$, where $\Lambda_{\text{max}}$ is the maximum Lyapunov exponent. We found out that the requirement $L\leq L_{\text max}=0.11$ is sufficient to obtain reliable results for the Lyapunov exponents $\Lambda_i$. 

\subsection{Asymptotic curve of rescaled Lyapunov spectrum}

\begin{figure}
\centering
\includegraphics[width=0.9\linewidth]{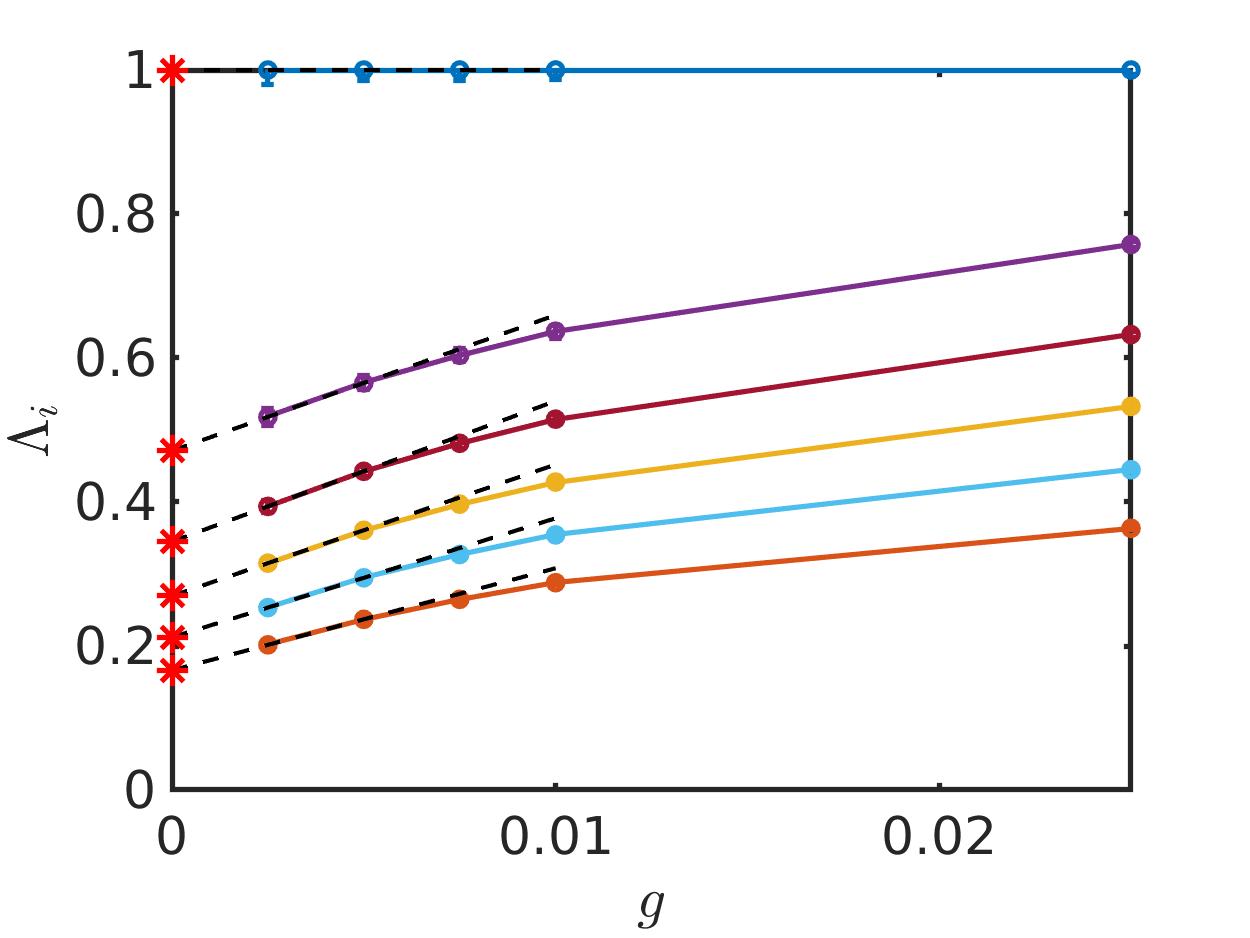}
\caption{The colored lines show $\Lambda_i$ versus $g$. The black dashed lines show the extrapolation process. The red stars show the points that are used in the plots of the asymptotic curves.}
\label{figS2}
\end{figure}

Fig.~\ref{figS2} depicts the extrapolation process used to obtain the results plotted in Fig. 5 of the main text. The values of the smallest and second smallest perturbation strength $g$ were determined based on the required accuracy $L\leq L_{\text max}$. To obtain information on the $\Lambda_i$ when $g$ approaches zero, we perform linear extrapolation of these two points. The red stars in Fig. \ref{figS2} exemplary indicate the values of the Lyapunov exponents (LEs) that are used in the plots of asymptotic curves resulting from such an extrapolation procedure. These LEs capture the behavior of the system as $g$ approaches zero and provide valuable insights into the system properties in the near-integrable regime.

\begin{figure}
\centering
\includegraphics[width=0.9\linewidth]{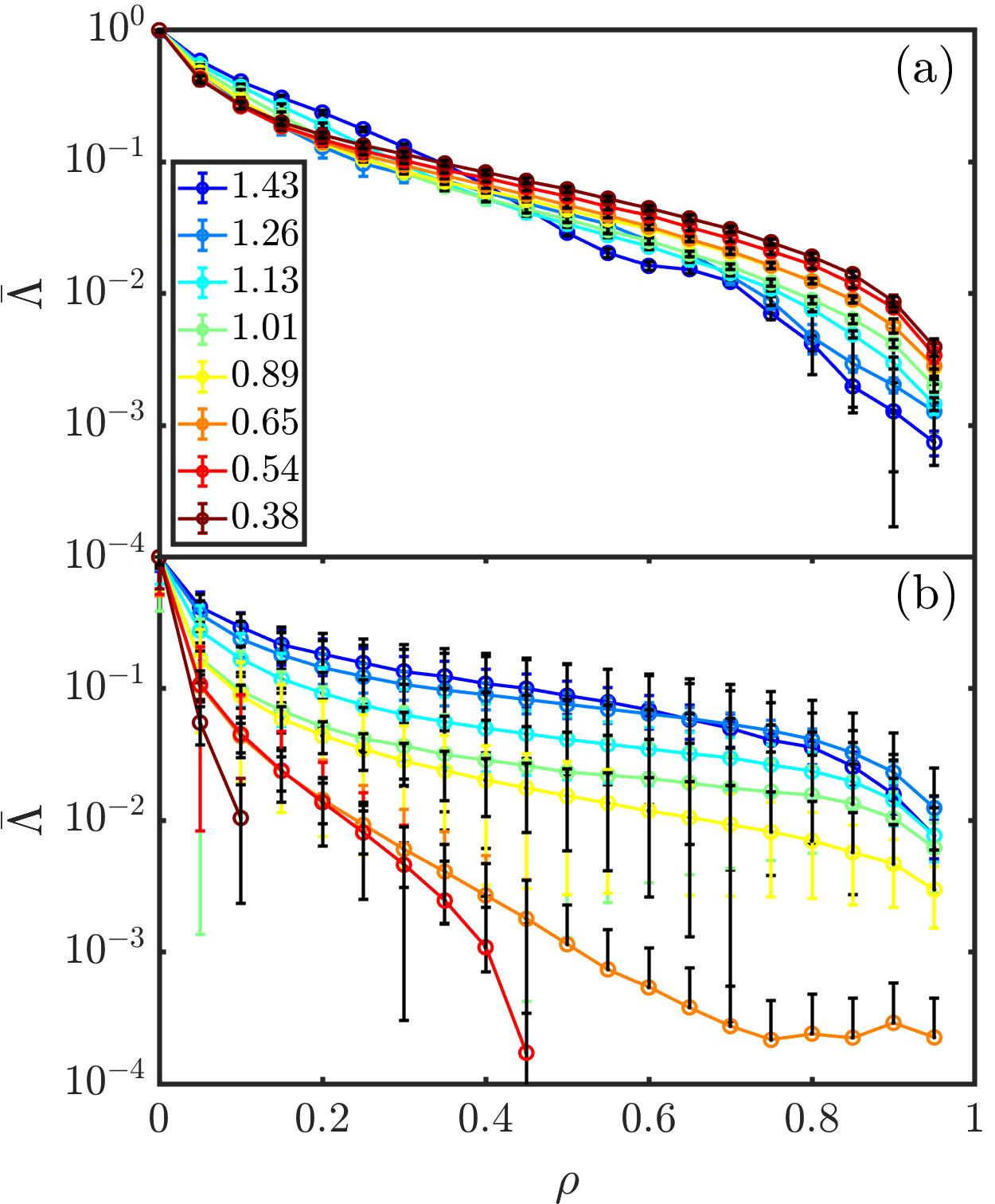}
\caption{
Asymptotic curves for rescaled LS versus rescaled index $\rho$ for the (a) ordered cases and (b) disordered cases in log scale. The angle $\theta$ varies from 1.43 (blue) to 0.38 (red).  
The error bars represent the time ($\sigma_t$ black) and ensemble ($\sigma_g$ colored) standard deviations. For all cases, the system size is $N = 200$. Unphysical error bars that would generate negative exponents are removed in panel (b).
}
\label{figS3}
\end{figure}

Using this extrapolation procedure, we have obtained the asymptotic curves for the rescaled Lyapunov spectra of both ordered and disordered cases, considering different values of $\theta$. These curves are depicted in Fig.~\ref{figS3}. The rescaled Lyapunov spectra provide valuable insights into the thermalization universality-class transition induced by disorder and the impact of localization on the system's behavior in the near-integrable regime.

\section{Fitting of Lyapunov Spectrum}
\begin{figure}
\centering
\includegraphics[width=0.9\linewidth]{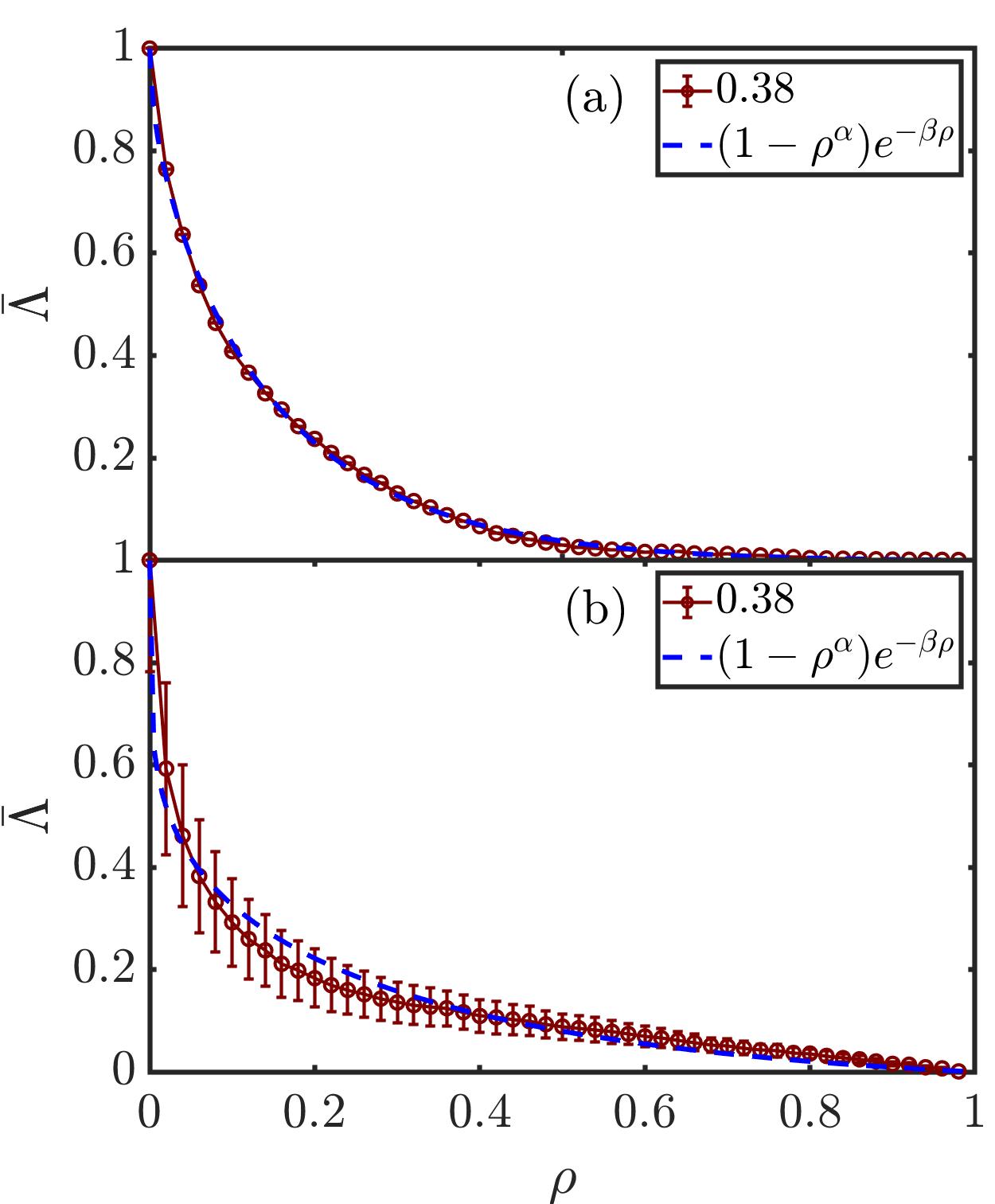}
\caption{
Asymptotic curves (red) for the rescaled LS versus rescaled index $\rho$ for the (a) ordered and (b) disordered cases in log scale. The angle $\theta$ is equal to 0.38.  
The error bars represent the ensemble ($\sigma_g$) standard deviations. 
The blue dashed curves are the fittings.}
\label{figS4}
\end{figure}
To further illustrate the distinct characteristics of the short- and long-range networks, we employ Eq.~(5) from the main text to fit the rescaled LS.

Fig.~\ref{figS4} presents one example for the fit of the proposed ansatz Eq. (5) of the main text to the rescaled LS for the ordered case and one for the disordered case. This fitting procedure allows us to better understand the behavior of the system and the underlying thermalization universality-class transition induced by Anderson localization. The fitted curves provide valuable insights into the relationship between the localization length $\xi$ and the thermalization behavior in the near-integrable regime.

\section{Fit Coefficient $\alpha$ for Asymptotic Curves of Ordered and Disordered Unitary Circuits Maps}

\begin{figure}
\centering
\includegraphics[width=0.9\linewidth]{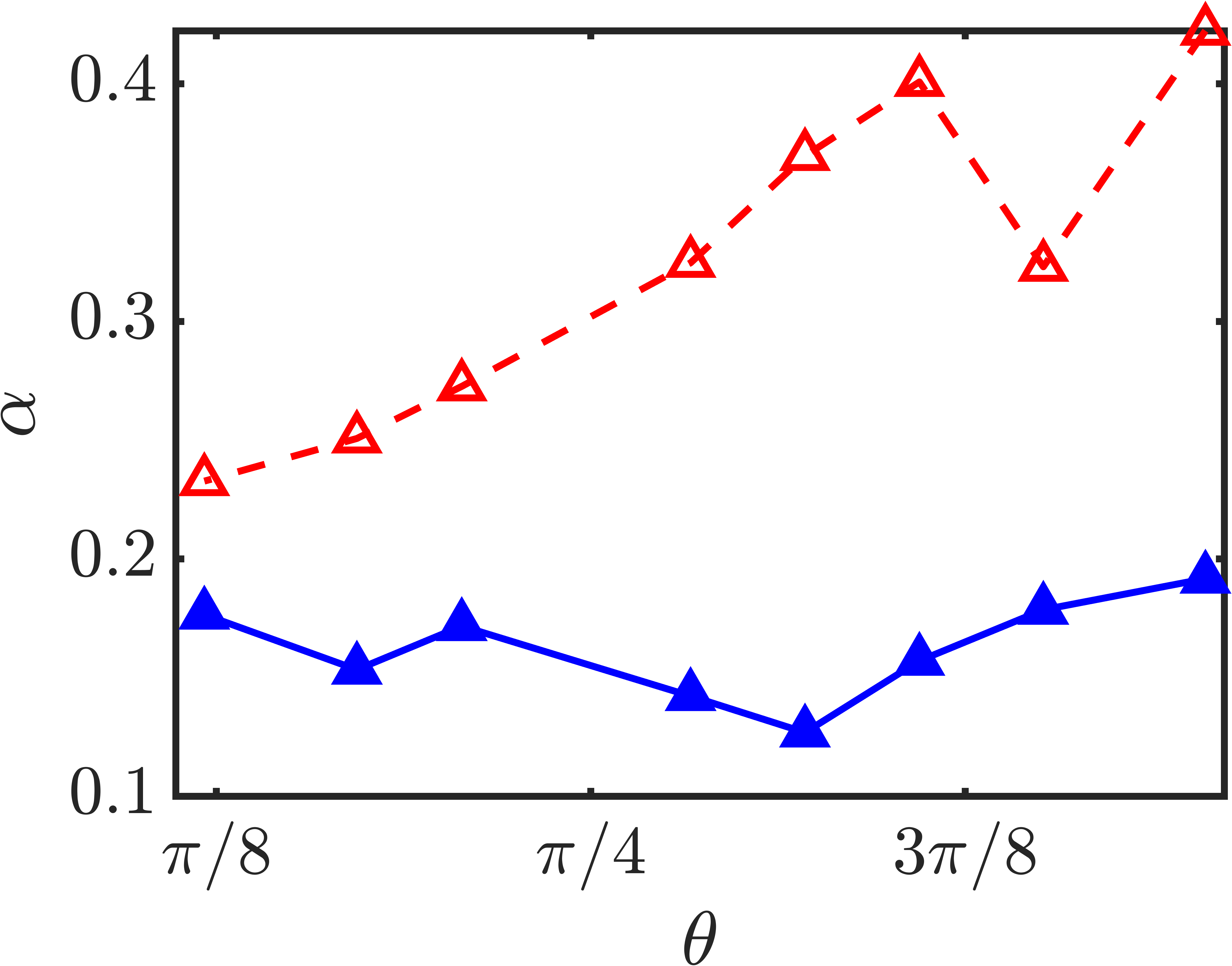}
\caption{Fit coefficient $\alpha$  ($\beta$ is shown in Fig.~5 of the main text) of Eq.~(5) in the main text versus $\theta$ for the asymptotic curves of the ordered (red empty triangles) and disordered (blue filled triangles) unitary circuits maps.}
\label{figS5}
\end{figure}

Fig.~\ref{figS5} displays the fit coefficients $\alpha$ obtained from Eq.~(5) of the main text as a function of $\theta$ for the asymptotic curves shown in Fig.~\ref{figS3}. In the plot, the red triangles connected by red dashed lines represent the fit coefficients for ordered systems, while the blue triangles connected by blue straight lines represent the fit coefficients for disordered systems.

\section{Variation of $\kappa_a$ with Different System Sizes}

\begin{figure}
\centering
\includegraphics[width=0.9\linewidth]{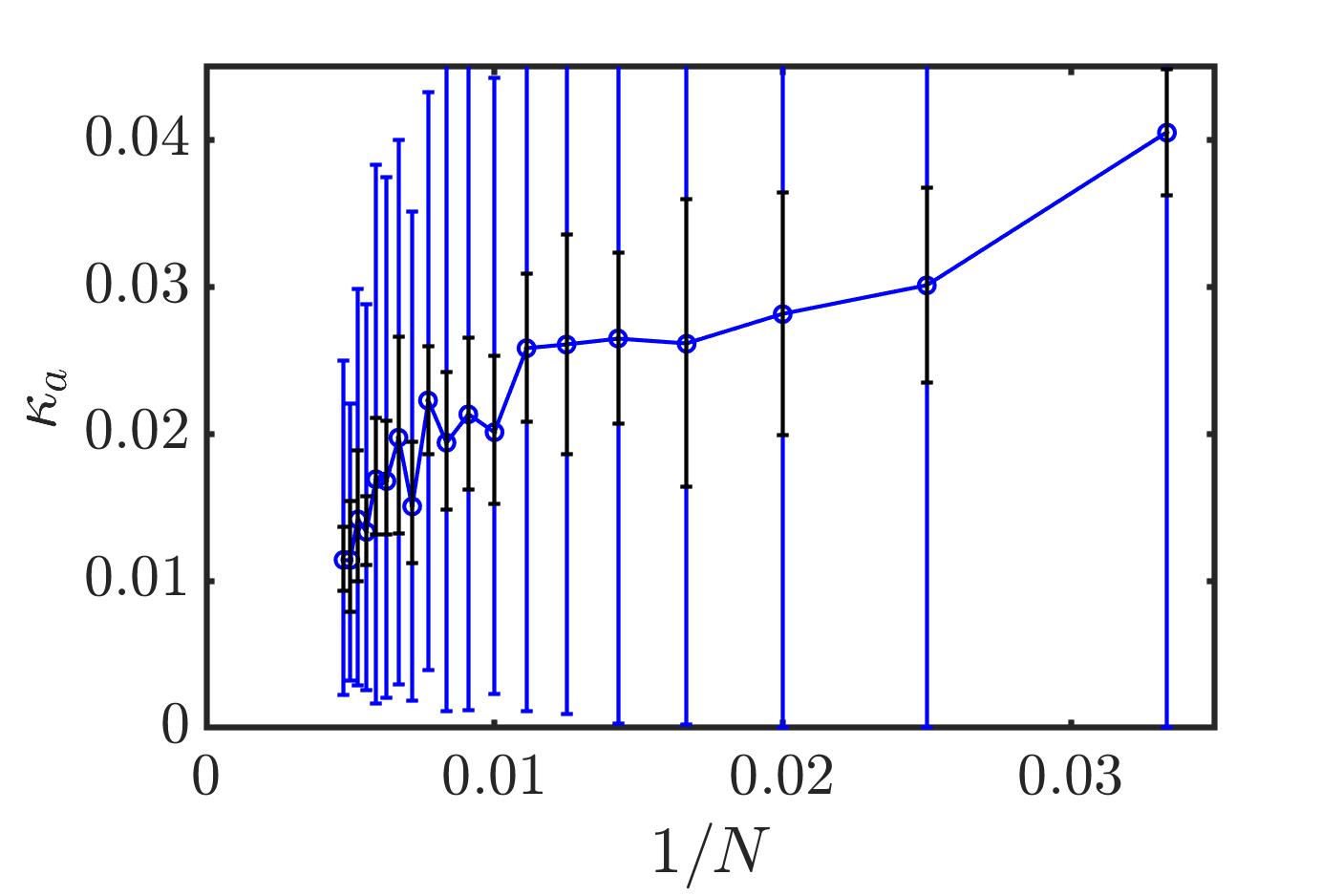}
\caption{
$\kappa_a$ versus $1/N$ with localization length $\xi=2$ $ (\theta=0.65)$ for disordered unitary circuits maps. The error bars represent the time ($\sigma_t$ black) and ensemble ($\sigma_g$ blue) standard deviations. The number of trajectories is 100.}
\label{figS6}
\end{figure}

In addition to inducing the transition from a long-range network to a short-range network by varying the localization length $\xi$, we can also achieve this transition by increasing the system size $N$. We have calculated the quantity $\kappa_a$ for different system sizes, all with the same localization length $\xi=2$, as presented in Fig.~\ref{figS6}.

We observe that as the system size $N$ increases, $\kappa_a$ decreases, providing further evidence for the long- to short-range network transition. Notably, the error bars deduced from ensemble averages are of comparable magnitude to the average value. This observation clearly indicates non-ergodicity and suggests that the maximum achievable time $T_s=10^8$ may not be sufficient for ergodization in the system. Note also that the ergodization time diverges much more rapidly in the short-range network regime \cite{danieli_dynamical_2019,mithun_dynamical_2019}, which will be explored in forthcoming works.

\end{appendix}

\bibliography{Reference}
\end{document}